\documentclass[a4paper,11pt]{article}
\pdfoutput=1 
\usepackage{jcappub} 
\usepackage[T1]{fontenc}

\usepackage{here}
\usepackage[utf8]{inputenc}
\usepackage{color}
\usepackage{amsfonts}
\usepackage{graphicx}
\usepackage{dsfont} 

\usepackage{bm}
\usepackage{graphicx}
\usepackage{amsmath}
\usepackage{amssymb}
\usepackage{enumitem}
\usepackage{hyperref}
\usepackage{subfigure}
\usepackage{booktabs}
\usepackage{array}
\usepackage[english]{babel}
\usepackage{tensor}
\usepackage{comment}
\usepackage[normalem]{ulem}
\allowdisplaybreaks

\newcommand{\WMAP}{\textsl{WMAP}}

\newcommand{\wmap}{{\WMAP}}
\newcommand{\Planck}{{\textsl{Planck}}}
\newcommand{\planck}{{\textsl{Planck}}}
\newcommand{\lcdm}{\ensuremath{\Lambda}CDM}

\newcommand{\kmsmpc}{\ensuremath{{\rm km\,s}^{-1}{\rm Mpc}^{-1}}}

\newcommand{\be}{\begin{equation}}
\newcommand{\ee}{\end{equation}}

\newcommand{\beq}{\begin{equation}}
\newcommand{\eeq}{\end{equation}}
\newcommand{\beqa}{\begin{eqnarray}}
\newcommand{\eeqa}{\end{eqnarray}}

\title{\boldmath Sound Horizon Independent Constraints on Early Dark Energy: The Role of Supernova Data}

\author[a,b,1]{Joshua A. Kable \note{Corresponding author.}}
\author[a,b]{Vivian Miranda}

\affiliation[a]{Department of Physics and Astronomy, Stony Brook University, Stony Brook, NY 11794, USA}
\affiliation[b]{C. N. Yang Institute for Theoretical Physics, Stony Brook University, Stony Brook, NY 11794, USA}

\emailAdd{joshua.kable@stonybrook.edu}
\emailAdd{vivian.miranda@stonybrook.edu}

\abstract{We assess the consistency of cosmological models that alter the size of the sound horizon at last scattering to resolve the Hubble tension with data from ACT + \planck\ CMB lensing, Big Bang Nucleosynthesis, and supernova data from Pantheon or Pantheon+. We use early dark energy (EDE) as an example model but conclude that the results could apply to other similar models. We constrain \lcdm\ and EDE with these data finding that while they can constrain \lcdm\ very tightly, EDE opens up the parameter space significantly and allows $H_0 > 72$ \kmsmpc. We combine these data with measurements from ACT + \planck\ TT650TEEE CMB primary anisotropy and galaxy baryon acoustic oscillations, and find that overall, EDE fits these data better than \lcdm\ at $\approx 2\sigma$. However, the fit to specifically the sound-horizon-independent measurements is worse for EDE than \lcdm. We assess this increase in $\chi^2$ coming from the sound-horizon-independent measurements and find that the best-fit model is still consistent with a random statistical fluctuation even with $H_0$ values around $72$ \kmsmpc. Finally, we find that supernova data play an important role in constraining EDE-like models with higher preferred values of $\Omega_m$, as preferred by Pantheon+, reducing the allowed parameter space for $H_0$ values greater than 70 \kmsmpc.  }

\begin{document}
\maketitle
\flushbottom

\section{Introduction}

In the last several years, increasingly precise measurements have led to constraints on cosmological parameters assuming the standard model of cosmology, \lcdm , that are in varying degrees of tension with direct measurements. The most widely discussed of these tensions is the Hubble tension or the disagreement in the present expansion rate of the universe, $H_0$, between early universe measurements assuming \lcdm\ and direct measurements in the late universe. This is usually described by the $4-6\sigma$ tension between the SH0ES cosmic distance ladder measurement of $H_0 \sim 73$ \kmsmpc\ \citep{riess/etal:2021b} and the \planck\ cosmic microwave background (CMB) measurement of $H_0 \sim 67$ \kmsmpc\ \citep{planck/6:2018}. However, this tension also persists across a number of other measurements, suggesting that this tension could be evidence for physics beyond \lcdm\ \citep[see e.g.][for a review]{Abdalla/etal:2022}. 

One complication in terms of resolving the Hubble tension is the measurement of the baryon acoustic oscillations (BAO) in the distribution of galaxies \citep{aylor/etal:2019,knox/millea:2020}. These BAO measurements are sensitive to the parameter combination $r_sH_0$, where $r_s$ is approximately the size of the sound horizon at the surface of last scattering\footnote{Galaxy BAO measurements are sensitive to the sound horizon during the drag epoch after recombination.}. Assuming \lcdm, the BAO data are consistent with either the SH0ES measurement (with lower values of $r_s$) or the \Planck\ measurement (with higher values of $r_s$), but not both at the same time \citep[see e.g. Figure 1 of ][]{knox/millea:2020}. 

One of the most popular extensions of the \lcdm\ model that attempts to resolve the Hubble tension is the early dark energy (EDE) model. It extends the standard flat \lcdm\ cosmology by including a new scalar field that becomes dynamical around matter-radiation equality \citep{Poulin/etal:2018}. This new scalar field increases the expansion rate of the universe relative to the \lcdm\ case and results in a lowering of the sound horizon preferred by CMB and BAO data, bringing them into better agreement with the SH0ES measurement. The key assumption of the EDE model is that the CMB, assuming \lcdm, prefers lower values of $H_0$ and higher values of $r_s$ because the sound horizon ruler is being miscalibrated as a result of the exclusion of this new scalar field component. 

There is a broad class of EDE scenarios that have been shown to alleviate the Hubble tension \citep[see e.g.][]{poulin/etal:2019,lin/etal:2019,Niedermann/etal:2019,karwal2021chameleon,sable/microphysics,McDonough:2021pdg,Lin:2022phm} when fit to \planck\ CMB anisotropy and gravitational lensing data \citep{planck/6:2018}, galaxy BAO data from BOSS DR 12 \citep{alam/etal:2017}, Type Ia supernova from the Pantheon compilation \citep{scolnic/etal:2018}, and a prior on $H_0$ from local measurements \citep{riess/etal:2021b}. For a review of the ability of EDE models to resolve the Hubble tension, see, for example, \cite{DiValentino:2021izs,Schoneberg:2021qvd}. While EDE models have been shown to alleviate the Hubble tension between \planck\ and SH0ES, these models do not fully resolve them. Moreover, the preference for $H_0 > 70$ \kmsmpc\ when fitting to \planck\ anisotropy data generally relies on the inclusion of a prior on $H_0$ from local measurements. Essentially, EDE models open up a degeneracy between the sound horizon and the Hubble constant, but there is not a significant preference to shift along this degeneracy without the prior on $H_0$. 

However, there has been some evidence of a preference for EDE models in primary CMB anisotropy data from the Atacama Cosmology Telescope (ACT), as well as the South Pole Telescope (SPT). In particular, \cite{hill2021atacama} found that there was a preference for EDE when fit to ACT data alone and in combination with a proxy for \wmap\ given by \planck\ TT with a multipole cut at $\ell = 650$. This was corroborated by \cite{Poulin/etal:2021} as well as \cite{Smith/etal:2022}, which also included SPT data and \planck\ TE and EE data. Similar conclusions were also found by \cite{Jiang/etal:2021,Jiang/etal:2022} who used a multipole cut at $\ell = 1000$ in TT. In all of these works, ACT or SPT data were shown to prefer $H_0 \sim 72-74$ \kmsmpc\ when assuming the EDE model even without including an $H_0$ prior. This contrasts the results when ACT and SPT data are fit assuming \lcdm, where the preferred $H_0$ constraints are comparable to the \planck\ equivalents \citep{ACTDR4,Dutcher/etal:2021}. 

Before proceeding, a few important points should be addressed. There is still no Bayesian preference for EDE in \planck\ temperature anisotropy data on small scales, though see \cite{Herold/etal:2022} for some evidence using a frequentist approach. Additionally, it is not clear whether the preference for EDE in ACT or SPT data is physical or simply the result of noisier measurements. More broadly, \citep{vagnozzi2023} argues that a solution to the Hubble tension likely cannot solely involve new physics in the early universe. It is nevertheless still interesting to study the ability of the EDE model to reconcile tensions between various data combinations both to probe the current cosmological tensions and to assess the robustness of the \lcdm\ model and current data. 

One novel way to probe the EDE scenario is to construct a combination of data whose measurements do not include information about the sound horizon. This combined data set should be less susceptible to the miscalibration of the size of the sound horizon proposed by EDE-like models. This has been explored in several works such as \citep{Pogosian/etal:2020,Philcox/etal:2022,ACTDR6MacCrann,ACTDR6Madhavacheril}. Recently, in \cite{ACTDR6Madhavacheril}, they constructed this type of data set, which included CMB lensing data from ACT and \planck, a prior on the baryon density given by $\Omega_bh^2 = 0.02233 \pm 0.00036$ from BBN constraints \citep{Mossa:2020gjc}, and a prior on $\Omega_m = 0.334 \pm 0.018$ corresponding to the constraint from Pantheon+ \citep{Brout/etal:2022}. Assuming the \lcdm\ model and some additional priors on cosmological parameters, \cite{ACTDR6Madhavacheril} found $H_0 = 65.0 \pm 3.2$ \kmsmpc\ when using ACT CMB lensing, and $H_0 = 64.9 \pm 2.8$ \kmsmpc\ when using both ACT and \planck\ CMB lensing. The latter constraint is $2.7\sigma$ lower than the SH0ES $H_0$ mean value.

In \cite{Philcox/etal:2022}, a sound-horizon-independent data combination, comprising \planck\ CMB lensing data, full shape galaxy power spectra, Pantheon+ supernova, and BBN constraints on the baryon density, was used to constrain \lcdm\ finding similar results to \cite{ACTDR6Madhavacheril}. They conclude that this disagreement between the constraints from their sound-horizon-independent data combination and SH0ES is either a result of an unknown systematic in at least one of the measurements or there needs to be an extension to the standard model of cosmology. In the latter case, an extended cosmological model that affects the size of the sound horizon at decoupling would importantly also have to affect the equality scale in a related way because these changes would not alter their equality scale based inferences of $H_0$.

Nevertheless, even the simplest EDE models do not solely affect the size of the sound horizon at recombination. For EDE to resolve the Hubble tension, it is necessary for the scalar-field to comprise approximately 10$\%$ of the energy budget of the universe around matter-radiation equality. This large of an early dark energy component has been shown to cause an enhanced early Integrated Sachs-Wolfe (eISW) effect that in turn results in a large increase in the cold dark matter density when fit to CMB anisotropy data \cite{Vagnozzi_2020}. Both the early dark energy field itself and the resulting shifts in the cold dark matter density can have non-negligible impacts on the equality scale that serves as the basis for the sound-horizon-independent measurements. It is therefore of interest to explore what constraints these data can place on EDE-like models.

To this point, \cite{hill2021atacama,Poulin/etal:2021,Smith/etal:2022} tested the impact of including \planck\ CMB lensing and Pantheon or Pantheon+ supernova constraints when fitting EDE and \lcdm\ to ACT primary anisotropy data. In general, the conclusion is that EDE fits this combined data better than \lcdm, though it fits \planck\ CMB lensing data worse. This point is consistent with results from \cite{hill/etal:2020,ivanov/etal:2020,D'Amico/etal:2020}, which have found that EDE tends to fit large-scale-structure (LSS) data worse than \lcdm\ because of the large increase in the cold dark matter density. In \citep{smith/etal:2023}, they directly assessed the consistency of EDE with sound horizon free measurements from BBN, full shape BOSS galaxy power spectra, \planck\ CMB lensing, and supernova from Pantheon+. They found that, assuming EDE, the preferred values of $H_0$ were not in tension with SH0ES. 

In this work, we continue these tests by examining how ACT CMB lensing can be used to constrain the EDE model alone and in conjunction with \planck\ CMB lensing. We also pay particular attention to the role that supernova data play, as there are non-negligible differences between constraints when using Pantheon, Pantheon+, or supernova from the Dark Energy Survey (DES). In Section~\ref{Theory}, we briefly summarize the EDE model theory. In Section~\ref{Methods}, we outline all of the data sets and combinations used in this paper as well as the numerical methods used in this work. In Section~\ref{Results}, we show the results of our tests of the EDE model using ACT CMB lensing data. Finally, in Section~\ref{Conclusions}, we offer conclusions and provide discussions. 

\section{Theory} \label{Theory}

This work aims to explore how sound-horizon-independent data, particularly from the latest ACT CMB lensing and supernova measurements, constrain extended cosmological models that predict changes to the size of the sound horizon before recombination. We focus on EDE specifically because it is the most explored example in the community. Thus, the results of this work can be more directly compared to other works. 

We use the EDE model discussed in \cite{Poulin/etal:2018,poulin/etal:2019}. In this scenario, an ultra-light axion scalar field, $\phi$, with a mass, $m$, becomes dynamical around matter-radiation equality. This scalar field has a potential given by 
\begin{equation}
    V\big(\phi\big) = m^2 f^2 \big(1 - \mathrm{cos}(\phi/f)\big)^n,
\end{equation}
where $f$ is the axion decay constant and $n$ sets the shape of the potential. We set $n = 3$, which has been shown to achieve good fits to cosmological data and find $H_0$ values that agree with SH0ES measurements when these models fit ACT and SPT data. 

The equations of motion at the background level are governed by the Klein-Gordon equation
\begin{equation}
\ddot{\phi} + 3H\dot{\phi} + \frac{dV}{d\phi} = 0,
\end{equation}
where $H$ is the Hubble parameter, and the dots refer to derivatives with respect to cosmic time. At early times, the Hubble friction term dominates and holds the scalar-field in place, while $V\big(\phi\big)$ sets the EDE energy density. However, the scalar field is eventually released from this Hubble friction as the universe expands, and it oscillates near the minimum of a potential that to first order goes as $V\big(\phi\big) \sim \phi^{2n}$. This leads to an effective equation of state parameter for the scalar field given by 
\begin{equation}
 w_n = \frac{n-1}{n+1},
\end{equation}
see \cite{poulin/etal:2019} for more details, though note that this approximation breaks down in the limit that the initial position of the scalar field $\theta_i \equiv \phi_i/f \sim 3$, where $\phi_i$ is the initial value of $\phi$. 

While the EDE theory is defined in terms of the scalar field parameters $m$ and $f$, it is often easier to think of the EDE model in terms of related phenomenological parameters. This is done for two reasons (1) the phenomenological parameters are easier to interpret in terms of the cosmology and (2) these parameters have been shown to converge more quickly using Markov Chains Monte Carlo (MCMC) than the theory parameters. 

We define the fraction of the total energy budget of the universe that is comprised of the early dark energy at a particular redshift, $z$, as $f_{\rm EDE}(z)$. We define the redshift at which this fraction achieves its peak value as $z_c$ and the fractional energy density as that redshift to be $f_{\rm EDE} \equiv f_{\rm EDE}(z_c)$. Finally, we use the $\theta_i$ parameter defined above meaning that EDE adds three new model parameters to the standard \lcdm\ scenario.

\section{Methods} \label{Methods}

\subsection{Observational Data} \label{Data}

We split the data used in this work into sound-horizon-dependent data sets and sound-horizon-independent data sets. In the following, we list all of the data that are used in various combinations in this work. Note that we do not use all of these data sets at once. We report all of the data set combinations used in this work in Table~\ref{tab:Labels}. For all data combinations, we assume a prior on $\tau = 0.06 \pm 0.01$\footnote{We use this prior on $\tau$ even for the sound-horizon-independent data combinations, which include no data sensitive to $\tau$. For this case, we tested the impact of fixing $\tau = 0.06$ compared to this prior choice and found no difference. We use the prior on $\tau$ for consistency with results in this paper when sound-horizon-dependent data are included}. For the sound-horizon-dependent measurements, we use

\begin{itemize}
    \item \textbf{ACT}: ACT DR4 TTTEEE power spectra using both the wide and deep patches \citep{ACTDR4}. We use \texttt{pyactlike}\footnote{https://github.com/ACTCollaboration/pyactlike}, and the \texttt{actpolite\_dr4} likelihood, which has been foreground marginalized. 
    \item \textbf{Planck}: Plik Lite 2018 TTTEEE likelihood \citep{planck/6:2018} with \Planck\ TT $ \ell \leq 30$ data. The Plik Lite likelihood is foreground marginalized. We do not include low $\ell$ EE data for consistency across all tests. 
    \item \textbf{BAO}: BAO data from the consensus BOSS DR 12 measurements \citep{alam/etal:2017}  in combination with the SDSS Main Galaxy sample \citep{ross/etal:2015}, and the 6dFGS survey \citep{beutler/etal:2011}. 
\end{itemize}
When including sound-horizon-dependent measurements, we combine ACT and \planck\ primary anisotropy data. We make a cut in TT for \planck\ at $\ell \leq 650$ to approximate combining ACT with \wmap\ data \citep{bennett/etal:2013}. We explore two cases for \planck\ primary anisotropy data, which are \planck\ TT650 and \planck\ TT650TEEE where in the latter case we include the full range TE and EE power spectra. For the sound-horizon-independent measurements, we use
\begin{itemize}
    \item \textbf{BBN Prior}: A prior on physical baryon density given by $0.02233 \pm 0.00036$ from Big Bang Nucleosynthesis (BBN) \citep{Mossa:2020gjc}.
    \item \textbf{Pantheon}: The full 2018 Pantheon Supernova catalog \citep{scolnic/etal:2018}.
    \item \textbf{Pantheon+}: A prior on $\Omega_m = 0.334 \pm 0.018$ from the updated Pantheon+ analysis \citep{Brout/etal:2022}.
    \item \textbf{DES}: A prior on $\Omega_m = 0.352 \pm 0.017$ from the DES supernova compilation \citep{DES/Supernova}.
    \item \textbf{ACT Lensing}: The ACT DR 6 CMB lensing potential power spectrum \citep{ACTDR6Madhavacheril}\footnote{https://github.com/ACTCollaboration/act\_dr6\_lenslike}. 
    \item \textbf{ACT + \planck\ Lensing}: The ACT DR 6 CMB lensing potential power spectrum combined with \planck\ CMB lensing \citep{ACTDR6Madhavacheril}. 
\end{itemize} 

\begin{table*}[!t]
\centering
\begin{tabular}{@{}cccc@{}}

 \multicolumn{2}{c}{} \\

\toprule

Labels  & Data Sets  \\ 

\toprule  
RS1 & ACT + \planck\ TT w/ $\ell_{\rm max}=650$ (a.k.a TT650) + BAO \\
RS2 & ACT + \planck\ TTTEEE w/ $\ell_{\rm max}^{\rm TT}=650$ (a.k.a. TT650TEEE) + BAO \\ 
P1 & ACT + \planck\ CMBL, priors on $A_s$ and $n_s$, and prior on $\Omega_bh^2$ from BBN  \\
P2 & P1 along with a cut in parameter space on $\Omega_ch^2 < 0.1446$ \\
NORS1 & ACT + \planck\ CMBL, prior on $\Omega_bh^2$ from BBN, and Pantheon SN data  \\
NORS2 & ACT + \planck\ CMBL, prior on $\Omega_bh^2$ from BBN, and Pantheon+ prior on $\Omega_m$  \\
NORS3 & ACT CMBL, prior on $\Omega_bh^2$ from BBN, and Pantheon SN data  \\
NORS4 & ACT CMBL, prior on $\Omega_bh^2$ from BBN, and Pantheon+ prior on $\Omega_m$  \\

\midrule

\end{tabular}
\caption{  Summary of the various data set labels used throughout the text. Here, CMBL refers to CMB lensing. For all data sets, we include a prior on $\tau = 0.06 \pm 0.01$. \label{tab:Labels}
}
\end{table*}

\subsection{Numerical Implementation} \label{Numerical}
To calculate cosmological observables to compare with measured data, we use a modified CAMB Boltzmann code \citep{Lewis_2000,Reboucas/etal:2024}. This code evolves the equations of motion at both the background and perturbation levels. To sample the posterior distributions for model parameters, we use MCMC samplers. To perform these MCMCs, we use the publicly available MCMC code \texttt{Cobaya} \citep{Cobaya}. To determine when the sampler has converged, we use a Gelman-Rubin convergence criteria of $R-1 = 0.02$ \citep{gelman/rubin:1992}. In many cases in this work, we perform statistical tests of chains that require greater convergence. For these cases, we use a convergence criteria of $R-1 = 0.005$. We use the BOBYQA algorithm built into \texttt{Cobaya} to find the best-fit set of parameter values. For the BOBYQA algorithm, we use the mean values from the chains as the starting point. As our default, we set the \texttt{CAMB} \texttt{AccuracyBoost = 1.5} and \texttt{CAMB} \texttt{LensPotentialAccuracy = 3}. In all cases where we include ACT CMB lensing data, we set \texttt{LensPotentialAccuracy = 4} as suggested by the ACT collaboration.

We sample the posteriors for both EDE and \lcdm. For the EDE-specific model parameters, we impose flat priors given by $1000 \leq z_c \leq 20000$, $0.001 \leq f_{\rm EDE} \leq 0.6$, and $0 \leq \theta_i \leq \pi$. These choices are consistent or broader than the same prior choices made in \cite{hill2021atacama}. For $z_c$, we impose the flat prior on $z_c$ instead of Log$_{10}(z_c)$. 

\section{Results} \label{Results}

\subsection{Sound-Horizon-Dependent Constraints} \label{SHDC}

\begin{figure*}[!tbp]
  \centering
 \includegraphics[width=6.0in, height=6.0in]{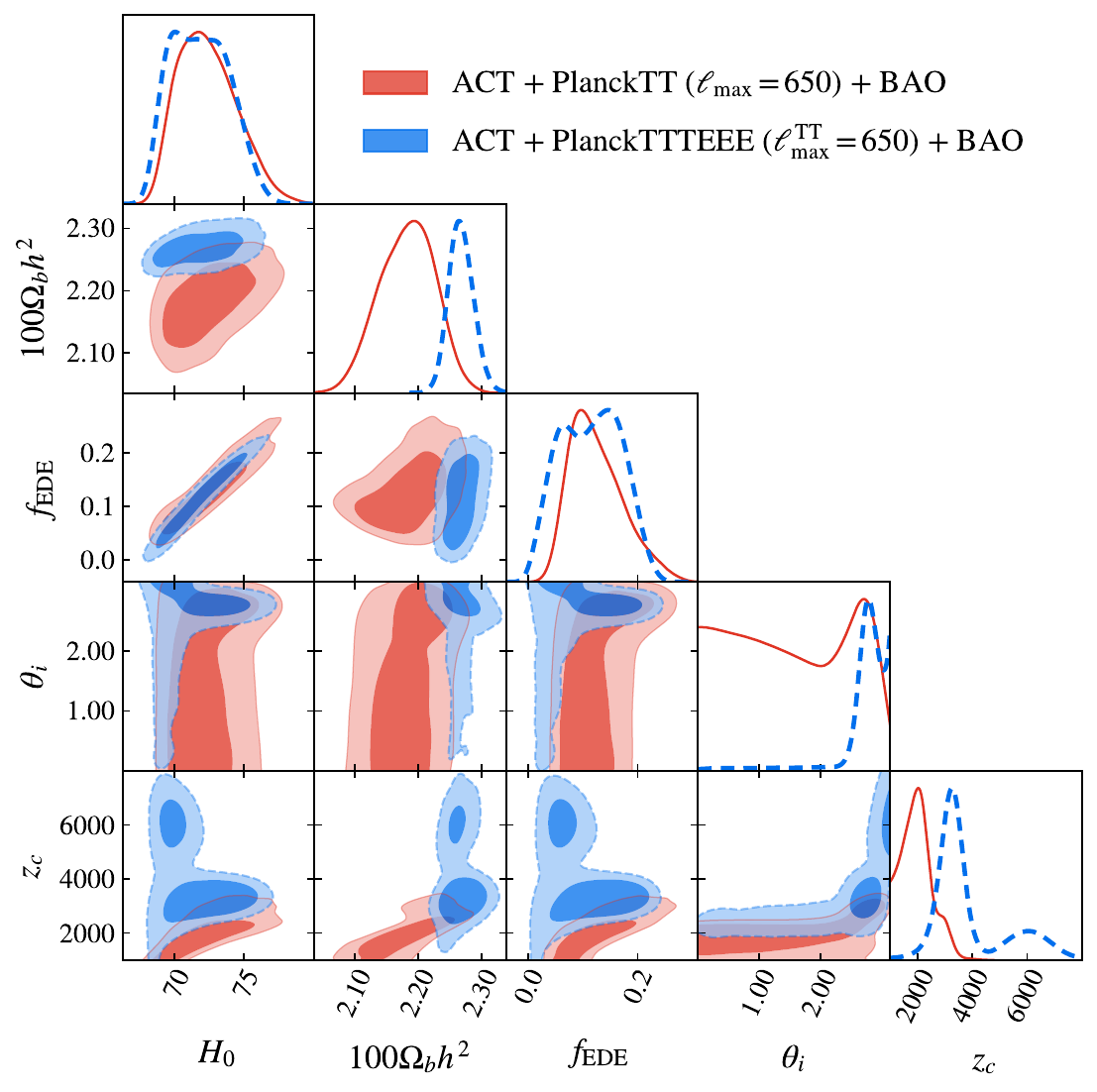}
   \vspace*{-0.4cm}
  \caption{ Plot of the posterior parameter distributions for EDE constrained by ACT + \planck\ TT650 + BAO (red) or ACT + \planck\ TT650TEEE + BAO (blue). Overall, we find good consistency with previous works on the behavior of EDE when fitting to these data combinations, including allowing for $H_0 $  values consistent with the SH0ES distance ladder measurements without including a prior on $H_0$. The inclusion of \planck\ TEEE data results in a shift upward and uncertainty reduction for $\Omega_bh^2$, the presence of a bimodality in the EDE model parameters between lower and higher values of $z_c$, and a reduction of the allowed parameter space in the $\theta_i-H_0$ plane so that the high $H_0$ values branch is restricted to $\theta_i \gtrsim 2.5$. The $\theta_i$ parameter sets the speed of sound of the scalar-field, which highlights the importance of the exact behavior of the speed of sound of the EDE scalar-field in allowing EDE to alleviate the Hubble tension when fit to these data. 
\label{fig:EDE_SH}}
\end{figure*}

In this subsection, we explore constraints from sound-horizon-dependent data on the \lcdm\ and EDE model parameters. We sample the posteriors for both models using two combinations of data, which are ACT + \planck\ TT650 + BAO (labeled RS1) and ACT + \planck\ TT650TEEE + BAO (labeled RS2). These data sets correspond to the choices used in \cite{hill2021atacama} and \cite{Smith/etal:2022} for primary CMB anisotropy data, respectively. We show the corresponding constraints in Table~\ref{tab:SH_constraints}. We additionally show the posteriors for select EDE parameters constrained by these two data sets in Figure~\ref{fig:EDE_SH}.

\begin{table*}[!t]
\setlength{\tabcolsep}{5pt}
\centering
\begin{tabular}{@{}cccccc@{}}

 \multicolumn{4}{c}{} \\

\toprule

 &  \lcdm: RS1   & \lcdm: RS2    & EDE: RS1  & EDE: RS2  \\ 
\toprule  

{\boldmath$100\theta_*$} & $1.04237\pm 0.00060$ & $1.04140^{+0.00040}_{-0.00031}$ &  $1.04122\pm 0.00070$& $1.04084^{+0.00050}_{-0.00071}$  \\
{\boldmath$\Omega_\mathrm{b} h^2$} & $0.02237\pm 0.00019$ & $0.02249^{+0.00017}_{-0.00011}$ & $0.02180^{+0.00084}_{-0.0011}$ & $0.02268^{+0.00018}_{-0.00021}$ \\
{\boldmath$\Omega_\mathrm{c} h^2$} & $0.1193^{+0.0012}_{-0.0017}$ & $0.11904^{+0.00079}_{-0.0017}$ & $0.1349^{+0.0061}_{-0.010}$ & $0.1312^{+0.0060}_{-0.0080}$ \\
{\boldmath$\log(10^{10} A_\mathrm{s})$} & $3.058\pm 0.021$ & $3.058^{+0.021}_{-0.017}$ & $3.073\pm 0.027$ & $3.077\pm 0.021$ \\
{\boldmath$n_\mathrm{s}   $} & $0.9761^{+0.0052}_{-0.0046}$ & $0.9743^{+0.0054}_{-0.0041}$ & $0.984^{+0.013}_{-0.017}$ & $0.9916\pm 0.0087$ \\
{\boldmath$\tau_\mathrm{reio}$} & $0.059\pm 0.010$ &  $0.0602^{+0.011}_{-0.0086} $ & $0.0565\pm 0.0097$ & $0.0603\pm 0.0095$ \\
{\boldmath$f_{\rm EDE}$} & --- & --- & $0.125^{+0.035}_{-0.061}$ & $0.113\pm 0.055$ \\
{\boldmath$z_c$} & --- & --- & $2000^{+400}_{-700}$ & 3900 [2600, 3800] \\
{\boldmath$\theta_i$} & --- & --- & $ < 2.20 \ (\textrm{at} \ 68 \%)$ & $> 2.73 \ (\textrm{at} \ 68 \%)                   $\\
\midrule

{\boldmath$H_0$} & $68.02^{+0.68}_{-0.52}$  & $67.90^{+0.75}_{-0.35}$ & $72.4^{+1.7}_{-2.6}$ & $71.8^{+1.9}_{-2.5}$ \\
{\boldmath$\Omega_\mathrm{m}$} & $0.3078^{+0.0066}_{-0.0097}$ & $0.3086^{+0.0043}_{-0.010}$ & $0.3003\pm 0.0078$   & $0.2995\pm 0.0073$ \\
{\boldmath$\sigma_8$} & $0.819\pm 0.011$ & $0.8161\pm 0.0089$ & $0.850\pm 0.029$ & $0.844\pm 0.016$ \\
{\boldmath$r_s$} & $144.61^{+0.41}_{-0.33}$ & $144.59^{+0.38}_{-0.21}$ &  $137.2^{+4.7}_{-2.9}$ & $138.3^{+3.7}_{-3.1}$ \\
\midrule

\end{tabular}
\caption{ Mean and 68$\%$ confidence intervals for \lcdm\ and EDE constraints from either ACT + \planck\ TT650 + BAO (RS1) or ACT + \planck\ TT650TEEE + BAO (RS2). In all cases, a prior of $\tau = 0.06 \pm 0.01 $ is included. \label{tab:SH_constraints}
}
\end{table*}

Table~\ref{tab:SH_constraints} shows that the EDE model constraints from both data sets result in increases in the preferred values of $H_0$ to be in better agreement with the SH0ES distance ladder measurement, and this effect is achieved by lowering the size of the sound horizon, $r_s$. Additionally, this increase in the mean value of $H_0$ is preferred by the RS1 and RS2 data sets without needing to include an $H_0$ prior. Both data sets also exclude $f_{\rm EDE} = 0$ at the 95$\%$ confidence level. These parameter shifts are compensated for by increases in the mean values of $n_s$ and $\Omega_ch^2$. 

While both the RS1 and RS2 data sets qualitatively find similar EDE behavior, there are several interesting differences resulting from including \planck\ TEEE. In particular, the parameter space for $\Omega_bh^2$ is shifted upward and reduced resulting in slightly better agreement with the BBN constraint, $\Omega_bh^2 = 0.02233 \pm 0.00036$, from \cite{Mossa:2020gjc}. In \cite{hill2021atacama} Tables II and IV, it is shown that the preference from ACT + \planck\ TT650 for lower values of $\Omega_bh^2$ is driven by the \planck\ TT650 data. Further, \cite{Smith/etal:2022} shows in Table III that the ACT + \planck\ TT650TEEE data prefer higher $\Omega_bh^2$ values consistent with our result. 

Additionally, when \planck\ TEEE data are included, the posteriors for the EDE model parameters, particularly $z_c$, are bimodal with essentially two main redshifts at which the EDE scalar-field achieves its peak energy density value. The $z_c < 4000$ mode allows for higher $H_0$ values similar to the case when \planck\ TEEE data are excluded, while the $z_c > 4000$ mode does not. This bimodality for $z_c$ has been observed in other works studying EDE, such as in Figure 6 of \cite{Smith/etal:2022} and Figure 4 of \cite{Karwal/etal:2024}. 

Finally, the allowed parameter space in the $\theta_i-H_0$ plane is significantly reduced when these data are included. Specifically, while $\theta_i$ is poorly constrained by both RS1 and RS2, the $\theta_i$ values where high $H_0$ values are allowed is restricted to $\theta_i \gtrsim 2.5$ when \planck\ TEEE data are included. This is also evident in the $68 \%$ confidence intervals in Table~\ref{tab:SH_constraints} where $\theta_i < 2.2$ when excluding \planck\ TEEE data while $\theta_i > 2.73$ when \planck\ TEEE data are included. This same behavior is also observed when comparing Figure 6 of \cite{Smith/etal:2022} to Figure 2 of \cite{hill2021atacama}. 

It can often be easier to think of EDE in terms of the fluid approximation, where EDE can be represented using a phenomenological equation of state, $w_{\rm EDE}$, instead of tracking the evolution of the scalar-field using the Klein-Gordon equation. We tested using the fluid approximation instead of the full scalar-field approach. While we found the same qualitative behavior for $f_{\rm EDE}$ and $H_0$ when constrained by ACT + \planck\ TT650 + BAO data, we found the preference for $H_0 \gtrsim 70$ \kmsmpc\ was excluded when \planck\ TEEE data were included. In \cite{Poulin/etal:2018}, they find that the fluid approximation for EDE breaks down in the limit that $\theta_i \rightarrow 3$, which is the limit where larger $H_0$ values are allowed. This highlights the importance of the $\theta_i$ parameter, which sets the speed of sound of the EDE scalar-field, in the fits to these data.

\subsection{Sound-Horizon-Independent Constraints} \label{NORSC}

In this subsection, we explore the constraining power of ACT and \planck\ CMB lensing, BBN, and supernova data on the \lcdm\ and EDE model parameters. For consistency with previous work, we follow \cite{ACTDR6Madhavacheril} in the construction of these data combinations which in turn follows \cite{Baxter_Sherwin/2021}. The CMB lensing data are sensitive to the scale of matter-radiation equality, but the constraint is degenerate with $A_s$. To break this degeneracy, we impose a prior on $10^9A_s = 2.19 \pm 0.09$ corresponding to the conservative prior choice in \cite{Baxter_Sherwin/2021} that is motivated by primary CMB anisotropy measurements. This prior choice is $1.5-2$ times larger than constraints for \lcdm\ and EDE on $A_s$ from CMB and BAO data (see Table~\ref{tab:SH_constraints}). Without this prior choice, we find the uncertainty on $H_0$ approximately doubles for the \lcdm\ case. Similarly, we impose flat priors given by $0.87 \leq n_s \leq 1.07$ and $0.014 \leq \Omega_bh^2 \leq 0.032$. 

The equality scale itself is sensitive to the parameter combination $\Omega_m^{0.6}h$. To break this degeneracy, we include supernova data from Pantheon, Pantheon+, or DES, which are sensitive to the expansion history of the universe. In a flat \lcdm\ universe, the only free cosmological parameter to constrain with supernova data is $\Omega_m$. This also holds true for the EDE model because the EDE scalar-field redshifts away faster than radiation and is thus negligible over the redshift ranges measured by these supernova experiments. Finally, we fix $\Sigma m_{\nu} = 0.06$ eV.

Any deviations in the late universe from \lcdm\ will affect the constraints from this data set. These changes could, for example, include a non-cosmological constant late-time dark energy that would weaken the constraint on $\Omega_m$ from supernova data. Because we are already extending the \lcdm\ scenario to include the EDE scalar-field, these late-time phenomenological changes are less interesting to additionally include as they would represent multiple new cosmological effects at very different cosmological epochs.

With these choices, we find constraints on the Hubble constant given by
\begin{equation}
    H_0 = 64.9 \pm 3.4 \ \mathrm{km \ s}^{-1}\mathrm{ \ Mpc}^{-1},
\end{equation}
for ACT and \planck\ CMB lensing with the Pantheon+ prior on $\Omega_m$. Additionally, including a prior on the baryon density from BBN results in 
\begin{equation}
    H_0 = 64.3 \pm 2.4 \ \mathrm{km \ s}^{-1}\mathrm{ \ Mpc}^{-1}.
\end{equation}

These $H_0$ constraints are both lower than that obtained from CMB primary anisotropy measurements assuming \lcdm, which results from the high value of $\Omega_m$ preferred by Pantheon+ relative to CMB primary anisotropy constraints. This feature was previously noted in \cite{Philcox/etal:2022}. If instead we use the full Pantheon supernova compilation, we obtain
\begin{equation}
    H_0 = 68.6 \pm 3.3 \ \mathrm{km \ s}^{-1}\mathrm{ \ Mpc}^{-1},
\end{equation}
where we have again used ACT and \planck\ CMB lensing and the BBN prior on the baryon density. 

\begin{figure}[!tbp]
  \centering
  \hspace*{+2cm}
 \includegraphics[width=3.75in, height=3.75in]{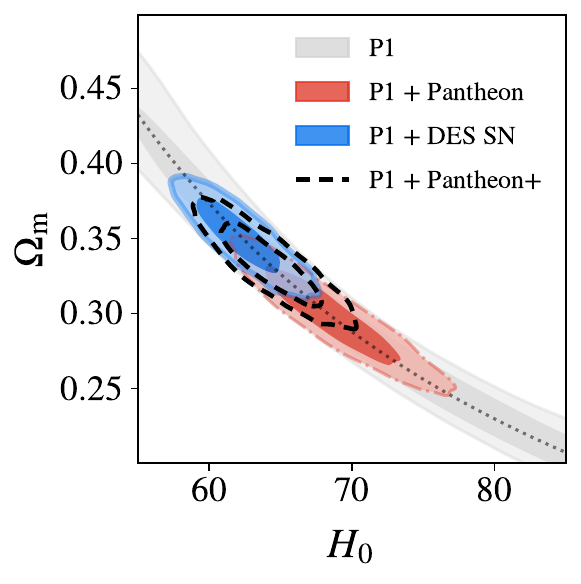}
  \hfill
     \vspace*{-0.4cm}
  \caption{Plot of the posterior parameter distributions for $\Omega_m$ and $H_0$ when \lcdm\ is constrained using various supernova data combined with data set P1, which includes ACT and \planck\ CMB lensing and priors on cosmological parameters outlined in the text. Within \lcdm, the ACT + \planck\ CMB lensing data constrain tightly the parameter combination $\Omega_m^{0.6}h$ (dotted black line). This degeneracy can be broken using supernova data, which in flat \lcdm, tightly constrain $\Omega_m$. For Pantheon+ and DES, we use priors on $\Omega_m$ resulting from constraints from supernova data. For the constraints on $H_0$, the choice of supernova data is very important. Pantheon+ and DES prefer relatively high values of $\Omega_m$ and thus prefer lower $H_0$ values when combined with P1.
\label{fig:LCDM_Omega_m_vs_H0}}
\end{figure}

We show the \lcdm\ model parameter posteriors when constrained by ACT and \planck\ CMB lensing, a BBN prior, and the priors on $A_s$ and $n_s$ above, which we label as data set P1, in Figure~\ref{fig:LCDM_Omega_m_vs_H0}. We additionally show the effect of including various different supernova data including Pantheon, Pantheon+, and DES. In the case of the latter two, we are using priors on $\Omega_m$, whereas for the former, we use the full Pantheon likelihood. In all cases, we find the posteriors are restricted to follow tightly the $\Omega_m^{0.6}h$ degeneracy, but the choice of supernova data used to break the degeneracy is important in determining both the preferred $H_0$ value and the uncertainty. In \cite{Baxter_Sherwin/2021}, they showed that tightening the uncertainty on $\Omega_m$ results in a marginal reduction in the uncertainty of $H_0$. However, the $H_0$ uncertainty is dependent on both the uncertainty of the $\Omega_m$ constraint from supernova data as well as the value of $\Omega_m$ as the $\Omega_m-H_0$ degeneracy is steeper at lower values of $H_0$. Hence, higher preferred values of $\Omega_m$ also tend to have lower uncertainties on $H_0$. 

\begin{table}[t]
\centering
\hspace{-1.0cm}
\begin{tabular}{@{}cccc@{}}

\multicolumn{2}{c}{} \\

\toprule

&  EDE  & EDE \\ 
&  P1+Pantheon  & P1+Pantheon+  \\ 
\toprule  

{\boldmath$100\theta_*$} & $1.029\pm 0.018$ & $1.012^{+0.018}_{-0.015}$ \\
{\boldmath$100 \Omega_\mathrm{b} h^2$} & $0.02234\pm 0.00036$ & $2.233\pm 0.036$ \\
{\boldmath$\Omega_\mathrm{c} h^2$} & $0.149^{+0.012}_{-0.036}$ & $0.145^{+0.011}_{-0.034}$ \\
{\boldmath$\log(10^{10} A_\mathrm{s})$} & $3.087\pm 0.040$ & $3.087\pm 0.041$ \\
{\boldmath$n_\mathrm{s}   $} & $0.961^{+0.041}_{-0.036}$ & $0.968^{+0.043}_{-0.035}$ \\
{\boldmath$\tau_\mathrm{reio}$} & $0.060\pm 0.010$ & $0.060\pm 0.010$ \\
{\boldmath$f_{\rm EDE}$} & $< 0.363$ \ (at 68$\%$)& $< 0.361$ \ (at 68$\%$)\\
{\boldmath$z_c$} & Unconstrained & $< 10400$ (at 68$\%$) \\
{\boldmath$\theta_i$} & Unconstrained &   Unconstrained    \\
\midrule
{\boldmath$H_0$} & $75.7^{+4.5}_{-9.0}$ & $70.9^{+3.5}_{-8.2}$ \\
{\boldmath$\Omega_\mathrm{m}$} & $0.298\pm 0.022$ & $0.332\pm 0.018$ \\
{\boldmath$\sigma_8$} & $0.825^{+0.026}_{-0.029}$ & $0.801\pm 0.024$ \\
{\boldmath$r_s$} & $131^{+10}_{-6}$ & $132^{+15}_{-7}$  \\
\midrule

\end{tabular}
\caption{Mean and 68$\%$ confidence intervals for EDE constraints from ACT + \planck\ CMB lensing and priors on select cosmological parameters (labeled P1) in combination with either Pantheon or Pantheon+ supernova data. In all cases a prior of $\tau = 0.06 \pm 0.01$ is included. \label{tab:EDE_NORS}
}
\end{table}

\begin{figure*}[!tbp]
  \centering
  \begin{minipage}[b]{0.38\textwidth}
 \includegraphics[width=3.1in, height=3.25in]{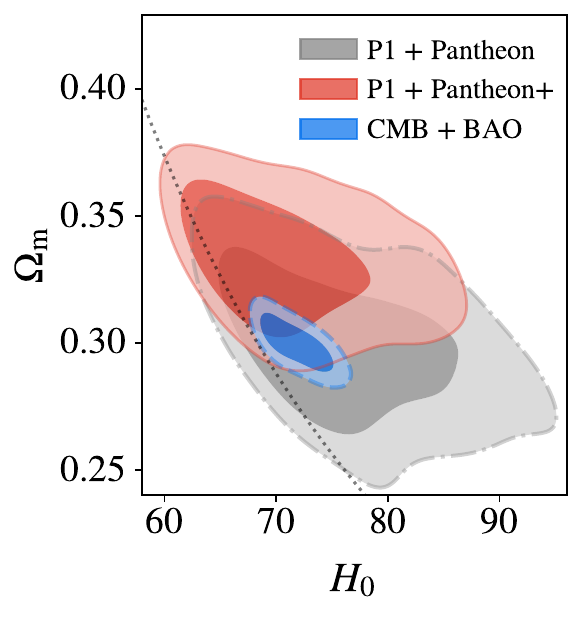}
  \end{minipage}
  \hfill
  \begin{minipage}[b]{0.38\textwidth}
    \hspace*{-1.8cm}
    \includegraphics[width=3.1in, height=3.25in]{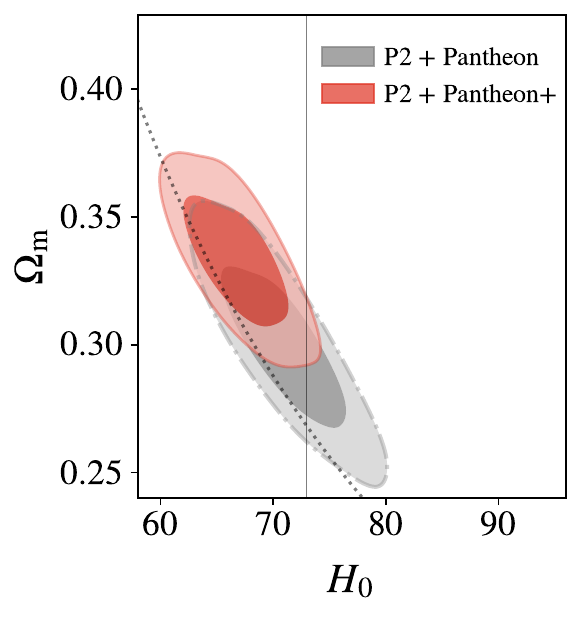}
  \end{minipage}
   \vspace*{-0.4cm}
  \caption{  (Left) Plot of the posterior distributions when EDE is constrained using supernova data from Pantheon (gray) or Pantheon+ (red) combined with data combination P1, which includes ACT and \planck\ CMB lensing data and priors on select cosmological parameters outlined in the text. We include the same \lcdm\ degeneracy curve as in Figure~\ref{fig:LCDM_Omega_m_vs_H0} for ease of comparison. EDE opens up parameter space and allows for higher values of $H_0$ primarily through the additional model parameter, $f_{\rm EDE}$. On their own, these combinations of data cannot rule out EDE as a model. Moreover, the EDE constraints from these data combinations are visually in agreement with the EDE constraints from ACT + \planck\ TT650TEEE + BAO  data (blue) and allow for higher $H_0$ values. (Right) Plot of the same EDE parameter posteriors as on the left from data combination P1 along with a cut on the cold dark matter density given by $\Omega_ch^2 < 0.1446$ (labeled P2).  Increasing $H_0$ without significantly altering $\Omega_m$ requires large shifts in $\Omega_ch^2$ to compensate. When these large values of $\Omega_ch^2$ are cut from the posterior distributions, the larger allowed values of $H_0$ are also removed as evidenced by the reduced fractions of both posteriors to the right of the SH0ES $H_0$ value (vertical black line). 
\label{fig:EDE_nors}}
\end{figure*}

In Table~\ref{tab:EDE_NORS} and Figure~\ref{fig:EDE_nors}, we show the EDE parameter constraints using data set P1 in combination with either Pantheon or Pantheon+. Figure~\ref{fig:EDE_nors} shows that in the low $f_{\rm EDE}$ limit, the EDE posteriors follow the same \lcdm\ degeneracy as expected as EDE reduces to \lcdm\ in this limit. However, the $f_{\rm EDE}$ parameter opens up a degeneracy with $H_0$ that these data are incapable of constraining on their own and thus allows for large increases in $H_0$ in the fits to both data combinations. While the mean value of $H_0$ for the P1 + Pantheon+ case is 70.9 \kmsmpc, the best-fit value from the chains is 74.39 \kmsmpc\ meaning that for both data combinations EDE can easily achieve $H_0$ values consistent with the SH0ES measurement. For comparison, in Figure~\ref{fig:EDE_nors} we show the $\Omega_m-H_0$ EDE constraints from ACT + \planck\ TT650TEEE + BAO (RS2 in Section~\ref{SHDC}), finding good agreement in this two dimensional parameter space, which suggests that combining these data sets together should result in consistency while achieving a relatively high value of $H_0$.

Table~\ref{tab:EDE_NORS} shows that the EDE specific parameters are mostly unconstrained by the data sets and priors. Additionally, the mean values of $100\theta_{*}$ for both cases have shifted by $1-1.5\sigma$ (relative to the uncertainties from these measurements) lower than the mean value preferred by CMB and BAO data with a significant increase in the uncertainties. We tested performing a parameter cut on these parameter posteriors such that we only keep chain points where $100\theta_{*} > 1.0388$. This corresponds to the lower bound of the 95$\%$ confidence interval in $\theta_*$ in the EDE fit to ACT + \planck\ TT650TEEE + BAO data (RS2). We find the resulting posterior still allows the same opening of the $\Omega_m-H_0$ degeneracy as without the parameter cut. 

In the second plot of Figure~\ref{fig:EDE_nors}, we show the same EDE posteriors but make a cut on $\Omega_ch^2$ so that we only keep chain points where $\Omega_ch^2 < 0.1446$. This threshold corresponds to a $2\sigma$ shift upward in $\Omega_ch^2$ from the EDE constraints from ACT + \planck\ TT650TEEE + BAO. With this cut, the allowed parameter space is significantly reduced to follow the \lcdm\ degeneracy curve and the higher values of $H_0$ are excluded. The allowed parameter space is reduced because the constraint on $\Omega_m$ means increasing $H_0$ necessarily must also increase $\Omega_mh^2$. Because of the BBN prior on $\Omega_bh^2$, the only parameter that can be adjusted to compensate for the large increases in $H_0$ is $\Omega_ch^2$. While the EDE fit to CMB and BAO data also prefers increases in the cold dark matter density that are large relative to the \lcdm\ constraints (see Table~\ref{tab:SH_constraints}), these increases are still smaller than those needed for the sound-horizon-independent data, particularly for Pantheon+, which prefers higher values of $\Omega_m$.

Table~\ref{tab:EDE_NORS} also shows that several of the constraints on the \lcdm\ parameters are prior dominated such as the constraints on $\Omega_bh^2$ and log(10$^{10}A_s$). Additionally, the constraints on $\Omega_m$ largely result from the supernova data or prior. Overall, these measurements cannot constrain the EDE phenomenology. A generic model that lowered the size of the sound horizon to increase $H_0$, could still be tightly constrained by these sound-horizon-independent data preventing resolution of the Hubble tension. If a cosmological model only lowered the size of the sound horizon at decoupling to increase $H_0$, then the $\Omega_m$ prior from the supernova data would result in an increase in the preferred value of $\Omega_mh^2$, which would in turn shift matter-radiation equality earlier allowing for more matter perturbation modes to grow for a longer time. This would in turn be disfavored by the CMB lensing data that is sensitive to the distribution of matter in the universe. Hence, even though the sound-horizon-independent data explored in this work are not directly sensitive to the physics of the sound horizon, they can still constrain changes to $H_0$ that result from shifts in the size of the sound horizon. For EDE, the lowering of the sound horizon is caused by a new dark energy component that slows the growth of matter perturbations. This approximately offsets the effects from the increase of $\Omega_mh^2$, so that given these current data, the EDE model is essentially unconstrained.

We note that using additional sound-horizon-independent measurements, such as including full shape BOSS galaxy power spectra as was explored in \citep{smith/etal:2023}, could provide more constraining power of the EDE model parameters. We leave this test to potential future analysis, though we expect this test will find similar results to \citep{smith/etal:2023}. In the following subsection, we study how combining ACT and \planck\ CMB lensing, supernova, and BBN data with the sound-horizon-dependent data (i.e. CMB and BAO) affects the constraints on EDE model parameters.

\subsection{Combining Sound-Horizon-Dependent and Independent Constraints} \label{sec:combined}

In the previous subsections, we explored the constraints on \lcdm\ and EDE model parameters from either sound-horizon-dependent (see Section~\ref{SHDC}) or sound-horizon-independent (see Section~\ref{NORSC}) data sets individually. We found that the sound-horizon-independent data, with appropriate prior choices, constrains the \lcdm\ model to tightly follow the $\Omega_m^{0.6}h$ degeneracy set by the matter-radiation equality scale. However, the EDE model is poorly constrained allowing for large increases in $H_0$ by opening up a new degeneracy direction with the $f_{\rm EDE}$. While these data cannot constrain EDE on their own, it is interesting to consider how these data can help constrain EDE when combined with sound-horizon-dependent data. 

In this subsection, we explore for the first time how including ACT + \planck\ CMB Lensing and supernova data from either Pantheon or Pantheon+ affect the various EDE parameter constraints when combined with CMB and BAO data. We take the chains from the \lcdm\ and EDE fits to ACT + \planck\ TT650TEEE + BAO (data combination RS2 from Section~\ref{SHDC}) and importance sample them using sound-horizon-independent data from ACT + \planck\ CMB lensing, a prior on the baryon density from BBN, and supernova data from either Pantheon or Pantheon+. We label these two cases as RS2 + NORS1 and RS2 + NORS2, respectively. Because we perform statistical tests of these data sets that require tight convergence, we use a convergence criteria of R-1 = 0.005 for the MCMC chains. 

We importance sample rather than run new MCMCs with both the sound-horizon-dependent and independent data together because it is possible that these two data combinations could be in tension with one another assuming a particular cosmological model. If there is good agreement between the two data sets given a cosmological model, then the combined MCMC will be equivalent to this procedure. However, if the two data sets are in tension, then all of the points sampled using the sound-horizon-dependent data sets will generate large $\chi^2$ values for the sound-horizon-independent data sets when importance sampled. In the case where the two data sets are in tension assuming a given cosmological model, running a full MCMC would result in a compromise between the two data sets based on the constraining power of each one. As a consistency check, we additionally ran full MCMCs corresponding to both the RS2 + NORS1 and RS2 + NORS2 cases, though to only R-1 = 0.05. We find good agreement between the two methods.

\begin{table*}[!htbp]
\small
\setlength{\tabcolsep}{3pt}
\centering
\resizebox{\columnwidth}{!}{%
\begin{tabular}{@{}cccccc@{}}
\multicolumn{4}{c}{} \\
\toprule
 &  \lcdm:    & \lcdm:   & EDE:  & EDE:   \\ 
 &  RS2   & RS2 + NORS1    & RS2  & RS2 + NORS1  \\ 
\toprule  
\vspace{-0.1cm}
{\boldmath$100\theta_*$} & $1.04140^{+0.00040}_{-0.00031}$   & $1.04144\pm 0.00033$  & $1.04084^{+0.00050}_{-0.00071}$  & $1.0411^{+0.0010}_{-0.0010}$  \\
& (1.04126) & (1.04125) & (1.04060) & (1.04066) \\
\vspace{-0.1cm}
{ \boldmath$100 \Omega_\mathrm{b} h^2$} & $2.249^{+0.017}_{-0.011}$ &  $2.250\pm 0.012$  & $2.268^{+0.018}_{-0.021}$  & $2.258\pm 0.016$  \\
& (0.02246) & (0.02244) & (0.02262) & (0.02253) \\
\vspace{-0.1cm}
{\boldmath$\Omega_\mathrm{c} h^2$} & $0.11904^{+0.00079}_{-0.0017}$ & $0.11876\pm 0.00088$ & $0.1312^{+0.0060}_{-0.0080}$  & $0.1277^{+0.0041}_{-0.0071}$ \\
& (0.11844) & (0.11847) & (0.1345) & (0.1321) \\
\vspace{-0.1cm}
{\boldmath$\log(10^{10} A_\mathrm{s})$} & $3.058^{+0.021}_{-0.017}$ & $3.067\pm 0.014$ & $3.077\pm 0.021$ & $3.075\pm 0.015$\\
 & (3.067) & (3.070) &  (3.087) &  (3.079)\\
\vspace{-0.1cm}
{\boldmath$n_\mathrm{s}$} & $0.9743^{+0.0054}_{-0.0041}$  & $0.9743\pm 0.0041$ & $0.9916\pm 0.0087$  & $0.9864\pm 0.0079$  \\
& (0.9713) & (0.9707) & (0.9916) & (0.9867) \\
\vspace{-0.1cm}
{\boldmath$\tau_\mathrm{reio}$} & $0.0602^{+0.011}_{-0.0086}$  & $0.0646\pm 0.0077$   & $0.0603\pm 0.0095$  & $0.0613\pm 0.0079$  \\
& (0.0606) & (0.0616) & (0.0589) & (0.0560) \\
\vspace{-0.1cm}
{\boldmath$f_{\rm EDE}$} & --- & --- & $0.113\pm 0.055$  & $0.085^{+0.042}_{-0.064}$  \\
&  &  & (0.147) & (0.128) \\
\vspace{-0.1cm}
{\boldmath$z_c$} & --- & --- & $3900 \ [2600,3800]$  & $3900^{+2000}_{-2000}$\\
 &  &  &  (3400) &  (3300)\\
\vspace{-0.1cm}
{\boldmath$\theta_i$} & --- & --- & $> 2.73$ \ (at 68$\%$)  & $> 2.75$ \ (at 68$\%$) \\
&  &  & (2.79) & (2.80) \\
\midrule
\vspace{-0.1cm}
{\boldmath$H_0$} & $67.90^{+0.75}_{-0.35}$ & $68.01\pm 0.38$  & $71.8^{+1.9}_{-2.5}$  & $70.8^{+1.3}_{-2.3}$ \\
&  (68.10) &  (68.07) & (72.98) &  (72.24) \\
\vspace{-0.1cm}
{\boldmath$\Omega_\mathrm{m}$} & $0.3086^{+0.0043}_{-0.010}$  & $0.3068\pm 0.0051$  & $0.2995\pm 0.0073$  & $0.3012\pm 0.0063$  \\
& (0.3052) & (0.3055) & (0.2963) & (0.2960) \\
{\boldmath$\sigma_8$} & $0.8161\pm 0.0089$ & $0.8188\pm 0.0055$  & $0.844\pm 0.016$  & $0.837^{+0.011}_{-0.014}$  \\
& (0.8167) &  (0.8183) &  (0.8524) &  (0.844) \\
\vspace{-0.1cm}
{\boldmath$r_s$} & $144.59^{+0.38}_{-0.21}$ & $144.66\pm 0.22$ & $138.3^{+3.7}_{-3.1}$  & $140.0^{+3.5}_{-2.2}$ \\
 &  (144.77) & (144.77) &  (136.6) &  (137.7) \\
\midrule
\vspace{-0.1cm}
{\boldmath$\chi^2_{\rm{ACT \ TTTEEE}}$} & 294.8 & 294.8 & 287.3 & 287.2 \\
\vspace{-0.1cm}
{\boldmath$\chi^2_{\rm{PL \ TT650TEEE}}$} & 443.8 & 443.8 & 439.1 & 439.7 \\
\vspace{-0.1cm}
{\boldmath$\chi^2_{\rm{PL \ TT \ }\ell \leq 30}$} & 22.3 & 22.4 & 21.0 & 21.4 \\
\vspace{-0.1cm}
{\boldmath$\chi^2_{\rm{BAO}}$} & 5.2 & 5.2 & 6.0 & 5.8 \\
\vspace{-0.1cm}
{\boldmath$\chi^2_{\rm{Lensing}}$} & --- & 19.7 & --- & 21.0 \\
\vspace{-0.1cm}
{\boldmath$\chi^2_{\rm{Pantheon}}$} & --- & 1034.8 & --- & 1034.7 \\
\vspace{-0.1cm}
{\boldmath$\chi^2_{\rm{BBN}}$} & --- & 0.1 & --- & 0.3 \\
\vspace{-0.1cm}
{\boldmath$\chi^2_{\rm{\tau}}$} & --- & 0.0 & --- & 0.2 \\
\vspace{-0.1cm}
{\boldmath$\chi^2_{\rm{Total}}$} & 766.1 & 1820.9 & 753.5 & 1810.3 \\
\end{tabular}
}
\vspace{-0.1cm}
\caption{Mean and 68$\%$ confidence intervals for \lcdm\ and EDE constraints from ACT + \planck\ TT650TEEE + BAO (RS2) and RS2 importance sampled using ACT + \planck\ CMB lensing, a BBN prior, and Pantheon supernova data. Parenthesis denote the best-fit value. Here PL refers to \planck. In all cases, a prior of $\tau = 0.06 \pm 0.01 $ is included. \label{tab:NORS1_IS}
}
\end{table*}

\begin{table*}[!htbp]
\setlength{\tabcolsep}{5pt}
\centering
\hspace{1.0cm}
\begin{tabular}{@{}ccc@{}}
 \multicolumn{2}{c}{} \\
\toprule
 & \lcdm: RS2 + NORS2   & EDE: RS2 + NORS2  \\ 
\toprule  
{\boldmath$100\theta_*$} &  $1.04140^{+0.00040}_{-0.00031}$ (1.04120) & $1.04144\pm 0.00033$ (1.04066)  \\
{\boldmath$100 \Omega_\mathrm{b} h^2$} &  $2.250\pm 0.012$ (0.02242) & $2.255\pm 0.016$ (0.02251) \\
{\boldmath$\Omega_\mathrm{c} h^2$} &  $0.11876\pm 0.00088$ (0.11890)  & $0.1263^{+0.0025}_{-0.0064}$ (0.1315)  \\
{\boldmath$\log(10^{10} A_\mathrm{s})$} &  $3.067\pm 0.014$ (3.067) &  $3.070\pm 0.015$ (3.077) \\
{\boldmath$n_\mathrm{s}$} & $0.9743\pm 0.0041$ (0.9700) & $0.9836\pm 0.0076$ (0.9850) \\
{\boldmath$\tau_\mathrm{reio}$} &  $0.0646\pm 0.0077$ (0.0596) & $0.0596\pm 0.0079$ (0.0556) \\
{\boldmath$f_{\rm EDE}$} & --- & $< 0.0831$  \ (at \ 68$\%$) (0.120) \\
{\boldmath$z_c$} & --- & $4200^{+2000}_{-2000}$ (3300) \\
{\boldmath$\theta_i$} & --- & $> 2.72$ \ (at \ 68$\%$) (2.80) \\
\midrule

{\boldmath$H_0$} & $68.01\pm 0.38$ (67.88) & $70.01^{+0.86}_{-2.0}$ (71.76) \\
{\boldmath$\Omega_\mathrm{m}$} & $0.3068\pm 0.0051$ (0.3081) & $0.3051\pm 0.0060$ (0.3004)\\
{\boldmath$\sigma_8$} & $0.8188\pm 0.0055$ (0.8183) & $0.8337^{+0.0091}_{-0.013}$ (0.8431) \\
{\boldmath$r_s$} & $144.66\pm 0.22$ (144.68) & $140.9^{+3.4}_{-1.2}$ (138.2) \\
\midrule
{\boldmath$\chi^2_{\rm{ACT \ TTTEEE}}$}  & 294.7 &  287.6 \\
{\boldmath$\chi^2_{\rm{PL \ TT650TEEE}}$}  & 443.9  & 439.9  \\
{\boldmath$\chi^2_{\rm{PL \ TT \ }\ell \leq 30}$}  & 22.5 & 21.5 \\
{\boldmath$\chi^2_{\rm{BAO}}$}  & 5.3  & 5.4 \\
{\boldmath$\chi^2_{\rm{Lensing}}$}  & 19.8 & 21.1 \\
{\boldmath$\chi^2_{\rm{Pantheon+}}$}  & 2.1  & 3.5  \\
{\boldmath$\chi^2_{\rm{BBN}}$}  & 0.1  & 0.3 \\
{\boldmath$\chi^2_{\rm{\tau}}$}  & 0.0 & 0.2 \\
{\boldmath$\chi^2_{\rm{Total}}$}  & 788.3  & 779.4 \\
\end{tabular}
\caption{ Mean and 68$\%$ confidence intervals for \lcdm\ and EDE constraints from \planck\ TT650TEEE + BAO (RS2) and RS2 importance sampled using ACT + \planck\ CMB lensing, a BBN prior, and Pantheon+ supernova data. Parenthesis denote the best-fit value. Here PL refers to \planck. In all cases, a prior of $\tau = 0.06 \pm 0.01 $ is included. \label{tab:NORS2_IS}
}
\end{table*}

We tested the impact of using ACT + \planck\ TT650 + BAO (data combination RS1 from Section~\ref{SHDC}) for the sound-horizon-dependent data set as well as the impact of using ACT CMB lensing or \planck\ CMB lensing instead of ACT + \planck\ CMB lensing. We find that our conclusions are robust to these choices. We choose to use the ACT + \planck\ TT650TEEE + BAO and ACT + \planck\ CMB lensing because these are the most constraining cases. 

\begin{figure*}[!tbp]
  \centering
  \begin{minipage}[b]{0.4\textwidth}
 \includegraphics[width=3.0in, height=3.25in]{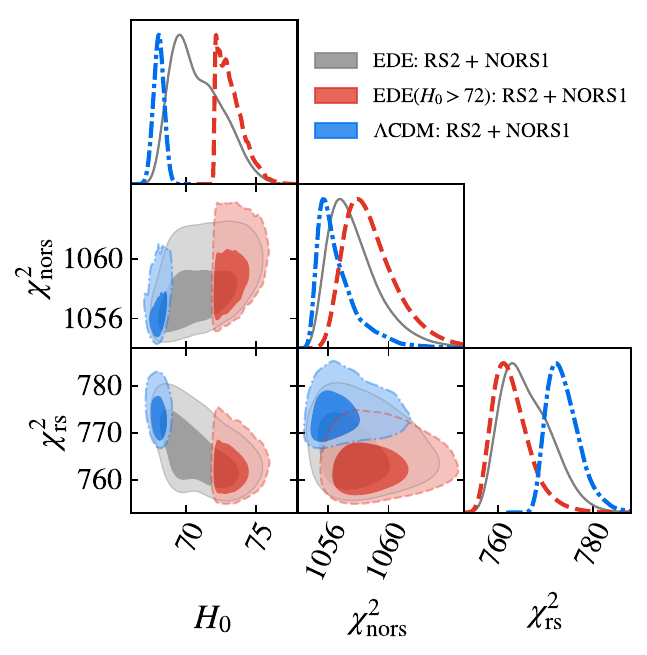}
  \end{minipage}
  \hfill
  \begin{minipage}[b]{0.4\textwidth}
    \hspace*{-1.4cm}
    \includegraphics[width=3.0in, height=3.25in]{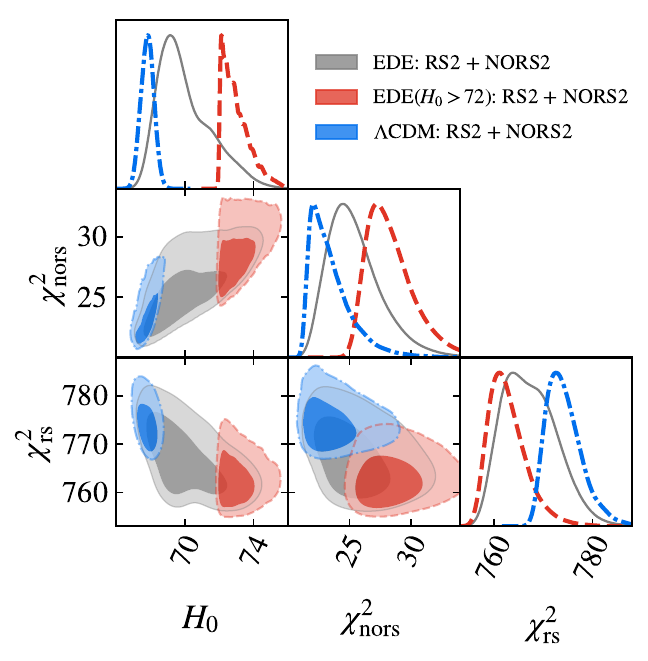}
  \end{minipage}
  \vspace{-0.4cm}
\caption{ Posterior distributions for \lcdm\ (blue) and EDE (gray) resulting from MCMCs constrained using ACT + \planck\ TT650TEEE + BAO (RS2) data that are importance sampled using data sets NORS1 or NORS2, which are comprised of ACT + \planck\ CMB lensing, BBN, and supernova data from either Pantheon (Left) or Pantheon+ (Right). For ease of comparison, we make a parameter cut for the EDE chains, $H_0 > 72$ \kmsmpc, and plot the resulting posteriors (red). For both plots, we find $\chi^2_{\rm{nors}}$ values are larger for EDE than for \lcdm\ suggesting EDE fits these data worse than \lcdm. However, in neither case is the worsening of the fit to the sound-horizon-independent data sufficient to rule out the EDE model. Using Pantheon data, the increase in $\chi^2_{\rm{nors}}$ is small relative to the reduction in $\chi^2_{\rm{rs}}$ even when allowing for high $H_0$ values.  This is particularly evident in the $\chi^2_{\rm{rs}}-\chi^2_{\rm{nors}}$ panel where a $\Delta \chi^2_{\rm{nors}}\sim 3$ between EDE and \lcdm\ is compared to a $\Delta \chi^2_{ \rm{rs}} \sim -12$. For the Pantheon+ case, the increase in $\chi_{rs}$ is larger because the prior on $\Omega_m$ prefers smaller values of $H_0$. In this case, the worsening of the $\chi^2_{\rm{nors}}~\sim 4$. 
\label{fig:chi2_post}}
\end{figure*}

We show the results of the importance sampling in Tables~\ref{tab:NORS1_IS} and \ref{tab:NORS2_IS} for the case with Pantheon supernova and the Pantheon+ prior respectively. We additionally ran an optimizer to find the best-fit set of points and report these values. In Figure~\ref{fig:chi2_post}, we break down the contributions to the $\chi^2$ into a component coming from sound-horizon-dependent data ($\chi^2_{\rm{rs}}$) and a component coming from sound-horizon-independent data ($\chi^2_{\rm{nors}}$). We additionally include a posterior corresponding to the same EDE chains with a cut in parameter space at $H_0 > 72$ \kmsmpc\ for visualization purposes. We do this because while it is interesting if EDE can fit all of the data sets used here better than \lcdm, ultimately, it is necessary for EDE to allow $H_0 > 72$ \kmsmpc\ to resolve the Hubble tension. 

Overall, the EDE model improves on the fit to both data combinations explored relative to \lcdm\ with a $\Delta \chi^2 = -10.6$ for NORS1 (including Pantheon) and $\Delta \chi^2 = -8.9$ for NORS2 (including Pantheon+). These reductions in $\chi^2$ correspond to Probability to Exceed (PTE) values assuming three degrees of freedom, for the three added EDE model parameters, of 0.014 (NORS1) and 0.03 (NORS2). These correspond to an overall $2.2\sigma$ and $1.9\sigma$ preference for EDE over \lcdm, respectively. 

Figure~\ref{fig:chi2_post} shows that EDE fits the ACT + \planck\ TT650TEEE + BAO data better than \lcdm. Moreover, as the value of $H_0$ increases within EDE, $\chi^2_{\rm{rs}}$ generally decreases. In contrast, $\chi^2_{\rm{nors}}$ tends to increase as $H_0$ increases. This gain is larger when Pantheon+ data are used than when Pantheon data are used. This is consistent with the constraints on $H_0$ in Tables~\ref{tab:NORS1_IS} and \ref{tab:NORS2_IS} where including the sound-horizon-independent data decreases the preferred value of $H_0$ relative to the cases with only the sound-horizon-dependent data. Nevertheless, we find that there is a region of parameter space where the EDE model has lower $\chi^2_{\rm{rs}}$ values and comparable $\chi^2_{\rm{nors}}$ values (within $\Delta \chi^2_{\rm{nors}} \sim 2-4$) even when the parameter space is restricted to only allow $H_0 > 72$ \kmsmpc. 

In both the RS2 + NORS1 and RS2 + NORS2 cases, the $\chi^2$ from the EDE fit to CMB lensing data from ACT and \planck\ is $\Delta \chi^2 = 1.3$ relative to \lcdm\ equivalents. While this does represent a worsening of the fit to the measured data, the ACT + \planck\ CMB lensing has 27 data points (18 for ACT and 9 for \planck) and 9 degrees of freedom. This means that the $\chi^2_{\rm{Lensing}} = 21.0 $ for the RS2 + NORS1 case corresponds to a PTE of 0.28, which suggests consistency between the CMB lensing data and the model. We perform this test more precisely in the following subsection where we include the same data sets and calculate the effective number of EDE parameters constrained by the sound-horizon-independent data given the constraints from the sound-horizon-dependent data, which are treated as a prior. 

In the best-fit cosmology case for RS2 + NORS1, the best-fit $H_0 = 72.24$ \kmsmpc, while the best-fit value for RS2 + NORS2 is $H_0 = 71.76$ \kmsmpc. These are shifts from the central values of the MCMC 68$\%$ constraints given by $H_0 = 70.8^{+1.3}_{-2.3}$ \kmsmpc\ and $H_0 = 70.01^{+0.86}_{-2.0}$ \kmsmpc\ for the NORS1 and NORS2 cases respectively. This means both of the best-fit $H_0$ values are outside of the 68$\%$ confidence intervals. The MCMC 95$\%$ confidence intervals are $H_0 = 70.8^{+3.5}_{-2.9}$ \kmsmpc\ and $H_0 = 70.01^{+3.3}_{-2.5}$ \kmsmpc, which suggests that there is a high degree of non-Gaussianity coming from the bimodal distribution for $z_c$ that explains this shift upward. The mean value is pulled toward lower $H_0$ values by the lower $H_0$ or higher $z_c$ branch, but the best-fit point only needs to find the minimum of the likelihood. While these best-fit $H_0$ values are not high enough to fully resolve the Hubble tension, a suitable model similar to EDE that assumes a miscalibration of the size of the sound horizon may be able to. 

While there is a 2.1$\sigma$ preference for a nonzero $f_{\rm{EDE}}$ when using only RS2 to constrain EDE, including NORS1 or NORS2 reduces this preference. For NORS1, the constraint is $f_{\rm{EDE}} = 0.085^{+0.042}_{-0.064}$ while for NORS2 $f_{\rm{EDE}} < 0.0831$ at 68$\%$ confidence level. Nevertheless, the 95$\%$ confidence level constraints are $f_{\rm{EDE}} < 0.170$ and $f_{\rm{EDE}} < 0.153$ suggesting that larger values of the EDE fraction of the energy budget of the universe are still allowed by these data. The EDE best-fit cosmology results in a $ \Delta \chi^2_{BBN} = 0.2$, relative to the \lcdm\ best-fit cases, for both the NORS1 and NORS2 cases. These increases are comparable to the shifts in $\chi^2$ when using alternative BBN data as was explored in \citep{Seto/etal:2021}, which primarily used \planck\ primary anisotropy and lensing data. 

These tests suggest that assuming the EDE model the sound-horizon-dependent and independent data are not in significant tension with each other, and importantly, the EDE model can still achieve $H_0$ values in good agreement with local measurements without the inclusion of an $H_0$ prior even with the latest CMB lensing data. We explore the level of this tension assuming \lcdm\ or EDE in more detail in the following subsection.

\subsection{QMAP: Goodness of Fit of the \lcdm\ and EDE Models with Sound-Horizon-Independent Data} \label{sec:QMAP}

In the previous subsection, we importance sampled MCMC chains resulting from sound-horizon-dependent data with sound-horizon-independent data. In particular, we used MCMCs of \lcdm\ and EDE constrained by ACT + \planck\ TT650TEEE + BAO data and importance sampled them using ACT + \planck\ CMB lensing, BBN, and either Pantheon (NORS1) or Pantheon+ (NORS2). We found the combined chains for EDE overall fit these data sets better than \lcdm\ with lower significance for the NORS2 case because of Pantheon+. This improvement comes mostly from the CMB primary anisotropy data as EDE fits to the CMB lensing data sets explored in this work are generally worse. This motivates checking whether this worsening of the fit to these data is statistically significant.

In this subsection, we quantify the goodness of fit of the \lcdm\ and EDE models to sound-horizon-independent data given a prior on parameter space resulting from the constraints from the sound-horizon-dependent data. Importantly, we are assessing the consistency of the models with the sound-horizon-independent data. We are using the sound-horizon-dependent constraints on parameters to restrict the allowed parameter space for the models. 

We do this using the $Q_{\rm MAP}$ statistic defined in \cite{raveri/hu:2019}. Here MAP refers to the maximum a posteriori point. The $Q_{\rm MAP}$ quantity is a goodness of fit statistic that is approximately a simple $\chi^2$ goodness of fit test, but with a proper accounting for the number of degrees of freedom based on the prior data included. It also evaluates the $\chi^2$ at the point in parameter space that optimally fits both the data and the prior. Quantitatively, 
\begin{equation}\label{eq:qmap}
    Q_{\rm{MAP}} = -2 \rm{ln} \mathcal{L}(\theta_p) - \rm{d}(2\pi) - \rm{ln}(|\Sigma|),
\end{equation}
where $\mathcal{L}(\theta_p)$ is the likelihood evaluated at the maximum a posteriori point, d is the number of data points in the data set, and $\Sigma$ is the data covariance matrix. 

Assuming a Gaussian prior, this reduces to 
\begin{equation}\label{eq:qmap}
    Q_{\rm{MAP}} \approx \chi^2(d - N_{\rm{eff}}),
\end{equation}
where $N_{\rm{eff}}$ is the effective number of parameters that are constrained by the data given that the prior already provides some constraints. Note that this $N_{\rm{eff}}$ does not refer to the number of relativistic degrees of freedom in the early universe. We adopt the notation of \cite{raveri/hu:2019}. 

In particular, 
\begin{equation} \label{eqn:Neff} 
N_{\rm{eff}} = N - \rm{tr}\big[\mathcal{C}^{-1}_{\Pi}\mathcal{C}_{p}\big],
\end{equation} 
where $N$ is the total number of model parameters, tr refers to the trace, $\mathcal{C}_{\Pi}$ is the parameter covariance matrix for the prior, and $\mathcal{C}_{p}$ is the parameter covariance of the posterior. In our case, $\mathcal{C}_{\Pi}$ is the parameter covariance matrix resulting from the constraints by ACT + \planck\ TT650TEEE + BAO data on \lcdm\ and EDE. Similarly, $\mathcal{C}_{p}$ is the parameter covariance matrix resulting from the importance sampling with the various NORS data sets. In the limit that the prior is completely uninformative, $N_{\rm{eff}}$ will reduce to the number of model parameters whereas in the limit where the prior provides all of the constraining power $N_{\rm{eff}} = 0$. In between these extremes, each parameter partially constrained by the prior will contribute a value to $N_{\rm{eff}}$ that is between 0 and 1. 

For each of the parameter values in Equation~\ref{eqn:Neff}, we follow the convention set by the publicly available Tensiometer code\footnote{https://github.com/mraveri/tensiometer/tree/master}. In particular, we regularize the eigenvalues of the $\mathcal{C}^{-1}_{\Pi} \mathcal{C}_{p}$ matrix by setting eigenvalues greater than one equal to one and eigenvalues less than 0 equal to 0. In our analysis, this is particularly important for the $z_c$ and $\theta_i$ parameters that are almost entirely unconstrained by the CMB lensing, BBN, and supernova data. 

To use the $Q_{\rm MAP}$ statistic, it is important to have good convergence of the parameter covariance matrix in order to calculate $N_{\rm{eff}}$. \cite{raveri/hu:2019} argue that it is therefore necessary for the MCMC to achieve $R-1 = 0.005$. Therefore, we use this convergence criterion for the MCMCs. For the EDE model, there are multiple parameters that have non-Gaussian posteriors when constrained by these data, which means that the corresponding parameter covariance matrices do not capture the full posterior behavior. Nevertheless, \cite{raveri/hu:2019} argues that this definition is sufficient as an approximation. 

For the importance sampling, we use four NORS data sets in this subsection. In all cases, we include a prior on $\Omega_bh^2$ from BBN. Additionally, the first two include ACT + \planck\ CMB lensing and either Pantheon (NORS1) or Pantheon+ (NORS2). The second two include ACT CMB lensing and either Pantheon (NORS3) or Pantheon+ (NORS4). For both the prior and posterior parameter matrices, we drop the rows and columns for $\tau$ as $\tau$ is mostly constrained by the prior. This is equivalent to marginalizing over all possible values of $\tau$. 

For the $Q_{\rm{MAP}}$ values, we need the likelihood evaluated at the maximum a posteriori point, which corresponds to the best-fit point given the data (sound-horizon-independent data) and the prior (the sound-horizon-dependent data). Hence, we use the best-fit values obtained through minimization for the combinations of RS2 and each of the NORS data sets for both \lcdm\ and EDE. We also tested using the best-fit values from the chains and found that our results are robust to this choice. Using these $Q_{\rm{MAP}}$ values as well as the number of degrees of freedom for each data combination (i.e. number of data points in the sound-horizon-independent data set - number of effective parameters constrained by the sound-horizon-independent data), we calculate PTE values.  

\begin{figure*}[!tbp]
  \centering
  \begin{minipage}[b]{0.45\textwidth}
  \hspace*{-0.12cm}
 \includegraphics[width=3.1in, height=2.3in]{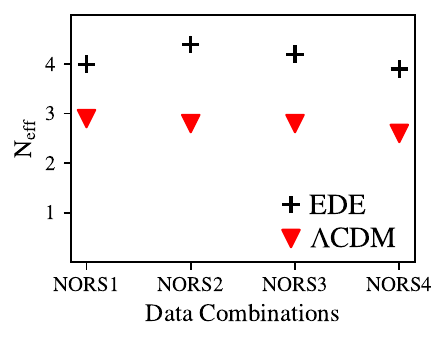}
  \end{minipage}
  \begin{minipage}[b]{0.45\textwidth}
  \hspace*{0.5cm}
 \includegraphics[width=3.1in, height=2.3in]{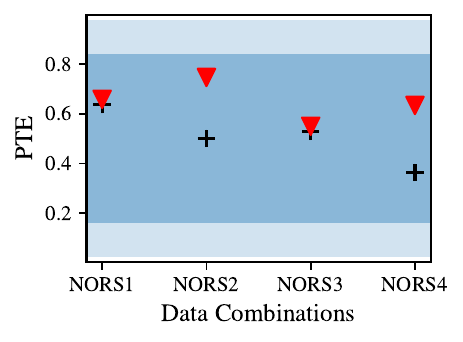}
  \end{minipage}
  \hfill
  \vspace{-0.4cm}
  \caption{ (Left) Effective number of parameters constrained by the corresponding sound-horizon-independent data sets given a prior set by the constraint on the corresponding model from ACT + \planck\ TT650TEEE + BAO (RS2) data. The first two data combinations include ACT + \planck\ CMB lensing + Pantheon (NORS1) or Pantheon+ (NORS2), while the second two include ACT CMB lensing + Pantheon (NORS3) or Pantheon+ (NORS4). These data constrain approximately three effective parameters for \lcdm\ and four effective parameters for EDE.  (Right) Plot of the PTE values resulting from minimization including both RS2 and the corresponding NORS data set given the number of NORS data points and the effective number of model parameters. The borders of the dark (light) blue band correspond to a $1\sigma$ (2$\sigma$) deviation from the expected value. For both models and for all cases the best-fit models are consistent with the data to within $1\sigma$. Including Pantheon+ results in a greater deviation between \lcdm\ and EDE PTE values than Pantheon partially because we use a prior on $\Omega_m$ that has only one degree of freedom compared to the 1048 degrees of freedom from Pantheon suggesting shifts in $\chi^2_{\rm{nors}}$ coming from CMB lensing data can be more easily absorbed in the Pantheon case. 
\label{fig:Neff_QMAP}}
\end{figure*}

In Figure~\ref{fig:Neff_QMAP}, we plot the $N_{\rm{eff}}$ and PTE values for each data combination and for both \lcdm\ and EDE. Each of the sound-horizon-independent data sets constrains about three parameters for \lcdm\ and four parameters for EDE, with the additional EDE parameter being primarily $f_{\rm{EDE}}$. There is some uncertainty to these numbers because of noise and non-Gaussianity. We made several tests to check the robustness of these results. For potential noise, we tested several amounts thinning the chains and calculating the new parameter covariance matrices for the prior and posteriors. For non-Gaussianity, we tested marginalizing over parameters by dropping their contributions from the parameter covariance matrices. We did one parameter at a time as well as for marginalizing over two parameters for $z_c$ and $\theta_i$, which are the least constrained parameters by the NORS data sets. In both cases, we find the $N_{\rm{eff}}$ values are robust to these choices. 

Overall, we find the PTE values for both \lcdm\ and EDE are consistent with being from a random draw to within one standard deviation for each of the data combinations. This is partially because while EDE does fit worse to the sound-horizon-independent data explored in this work than \lcdm, the \lcdm\ fits are slightly too consistent, so this worsening still results in all model and data combinations being consistent. For EDE, this means that even though EDE tends to fit the CMB lensing, BBN, and supernova data worse than \lcdm\ when restricted to the parameter space allowed by the CMB primary anisotropy and BAO data, the worse fit is still consistent with being a statistical fluctuation. 

Both cases that include Pantheon+ data have wider separations between the PTE values for \lcdm\ and EDE than the corresponding values when Pantheon is used. This occurs for two reasons. The first is simply the higher $H_0$ values generate larger $\chi^2$ from Pantheon+ than from Pantheon because Pantheon+ prefers a higher value of $\Omega_m$. The second is an artifact of the test itself. The Pantheon+ prior has only one degree of freedom, while the Pantheon data have 1048 from each of supernova. A difference in $\chi^2$ of a couple can be absorbed in the latter case, whereas it leads to a larger difference in PTE values in the former case. While this could motivate calculating PTE values for only the CMB lensing data sets, the cases with Pantheon+ find no significant deviation, and the EDE model fits the CMB lensing data sets approximately the same for the different supernova choices.

Both models tend to fit the data combinations including ACT CMB lensing data, instead of ACT + \planck\ CMB lensing data, slightly worse. This suggests that both models fit the \planck\ CMB lensing data included in the ACT + \planck\ CMB lensing likelihood equally well. This is not equivalent to the \planck\ CMB lensing data on their own as parts of the \planck\ CMB lensing likelihood used in combination with ACT are removed because there is cross-correlation. Additionally, the lowest EDE best-fit $H_0$ value is $H_0 = 71.76$ \kmsmpc\ from the RS2 + NORS2 combination, and the highest EDE best-fit value of $H_0 = 72.58$ \kmsmpc\ from the RS2 + NORS3 combination. This is important because EDE was proposed as a solution to the Hubble tension. The EDE model is consistent with the sound-horizon-independent data when limited by the sound-horizon-dependent parameter constraints, a better fit to the sound-horizon-dependent data, and allows for $H_0\sim72$ \kmsmpc. 

\subsection{Effect of Varying $H_0$}

\begin{figure*}[!thp]
  \centering
  \begin{minipage}[b]{0.45\textwidth}
  \hspace*{-0.2cm}
 \includegraphics[width=3.1in, height=3.1in]{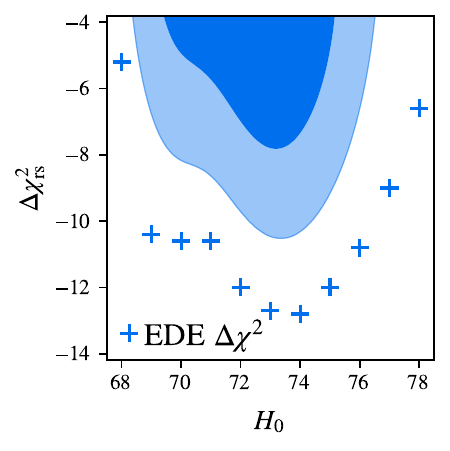}
  \end{minipage}
  \begin{minipage}[b]{0.45\textwidth}
  \hspace*{0.5cm}
 \includegraphics[width=3.1in, height=3.1in]{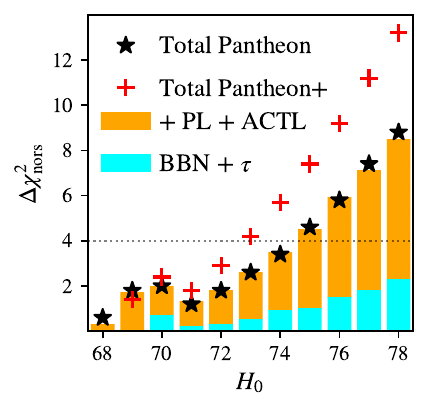}
  \end{minipage}
  \hfill
  \vspace{-0.4cm}
  \caption{ (Left) Plot of the difference in $\chi^2_{\rm{rs}}$ between EDE where $H_0$ is fixed to values between $H_0 = 68-78$ \kmsmpc, and \lcdm\ where $H_0$ is free. In both cases, the models are fit to ACT + \planck\ TT650TEEE + BAO data. We additionally include the MCMC posterior for the EDE fit to the same data with the best-fit \lcdm\ $\chi^2_{\rm{rs}}$ value subtracted. The largest difference in $\chi^2_{\rm{rs}}$ occurs when $H_0$ is fixed to 74 \kmsmpc. There is a plateau in the range $69-71$ \kmsmpc\ resulting from the bimodal feature in $z_c$. Using each of these best-fit cosmology values for both the \lcdm\ and EDE models, we evaluate the CMB lensing, BBN, and supernova likelihoods from NORS1 and NORS2. (Right) We plot the difference in $\chi^2_{\rm{nors}}$ for each EDE fixed $H_0$ value relative to the \lcdm\ case where $H_0$ is free to vary. Additionally, we break down the contributions to this $\chi^2$ from each of the individual likelihoods. As $H_0$ increases, the $\chi^2_{\rm{nors}}$ value tends to increase highlighting the difference with primary CMB anisotropy and BAO data. For both NORS1 and NORS2, this is because increasing $H_0$ tends to worsen the fit to the CMB lensing data. For NORS2, the Pantheon+ high preferred value of $\Omega_m$ also tends to worsen the fit, while Pantheon is largely unaffected. Nevertheless, even when using best-fit cosmologies for sound-horizon-dependent data, the total increase in $\chi^2_{\rm{nors}}$ is restricted to be smaller than about four up to $H_0 \sim 73$ \kmsmpc\ for Pantheon+ and $H_0 \sim 74$ \kmsmpc\ for Pantheon. 
\label{fig:H0_effect}}
\end{figure*}

In the previous subsection, we showed that there is generally good agreement between the EDE model and the sound-horizon-independent data explored given a prior on parameters from the sound-horizon-dependent data. In this subsection we explore how varying the value of $H_0$ changes the fit to the sound-horizon-independent data. In particular, we run a series of minimizations of the EDE fit to ACT + \planck\ TT650TEEE + BAO (i.e., RS2) data where we fix the value of $H_0$ to specific values between 68 and 78 \kmsmpc\ in 1 \kmsmpc\ intervals. 

For the start values of the minimizations, we take the MCMC chain and make cuts in $H_0$ on the parameter posteriors so that only points within 1 \kmsmpc\ of the fixed $H_0$ value are allowed. Using these cut chains, we find the averages of each of the model parameters and use these as the starting values for the minimization. For example, for the $H_0 = 72$ \kmsmpc\  test, we make cuts in parameter space so that we only include chain steps where $71 \leq H_0 \leq 73$ \kmsmpc, and then we find the averages of all model parameters for the remaining chain.

For comparison with \lcdm, we also run a single \lcdm\ minimization to the same data where $H_0$ is free to vary. We calculate the difference between the overall best-fit \lcdm\ $\chi^2_{\rm{rs}}$ value and the $\chi^2_{\rm{rs}}$ value for each fixed $H_0$ EDE best-fit cosmology. We plot these differences against the fixed $H_0$ in Figure~\ref{fig:H0_effect}. We find that in each fixed $H_0$ case, the EDE fit to the data is better (i.e. a negative $\Delta \chi^2_{\rm{rs}}$ value). The largest improvement in $\chi^2_{\rm{rs}}$ occurs at $H_0 = 74$ \kmsmpc\ where $\Delta \chi^2_{\rm{rs}} = -12.7$. This improvement in $\chi^2_{\rm{rs}}$ is only for the CMB anisotropy and BAO data and does not include an $H_0$ prior from local measurements.

There is a plateau in the plot of $\Delta \chi^2_{\rm{rs}}$ vs $H_0$ near $H_0 \sim 69-71$ \kmsmpc. For reference, we also show the posterior from the corresponding MCMC, which also appears to have this plateau feature in this region suggesting this is not a minimization error. The $z_c$ values for the $H_0 = {69,70}$ \kmsmpc\ best-fit cosmologies are order $6000$, while for the ${71,72}$ \kmsmpc\ best-fit cosmologies, the $z_c$ values are order $3000$. From Figure~\ref{fig:EDE_SH}, we deduce this plateau results from the bimodal distribution for $z_c$. 

Using these best-fit parameter values for the single \lcdm\ and each EDE minimization, we evaluate the likelihood given by priors on the optical depth and baryon density, CMB lensing data from ACT and \planck, and either Pantheon or Pantheon+ supernova data. We plot the difference in $\chi^2_{\rm{nors}}$ between the single \lcdm\ minimization and each EDE minimization against the fixed $H_0$ values in Figure~\ref{fig:H0_effect}. For $H_0 > 71$ \kmsmpc, there is a clear increase in $\Delta \chi^2_{\rm{nors}}$ meaning EDE cosmologies that prefer higher values of $H_0$ and optimally fit the sound-horizon-dependent data are disfavored by these sound-horizon-independent data. 

While EDE mostly fits the Pantheon data equally as well as \lcdm\ even as $H_0$ increases, EDE fits the Pantheon+ prior worse as $H_0$ increases. This prior may break down as a good approximation for the Pantheon+ constraints as $H_0$ increases because this limit corresponds to taking differences in the $\chi^2$ values calculated in the tails of an assumed Gaussian distribution. Nevertheless, the $\Delta \chi^2_{\rm{nors}} \lessapprox 4$ up to $H_0\sim 74$ \kmsmpc\ for Pantheon and up to $H_0\sim 73$ \kmsmpc\ for Pantheon+. The threshold of $\Delta \chi^2_{\rm{nors}} \sim 4$ corresponds to approximately a $2\sigma$ shift. 

One important note for this test is that these $\Delta \chi^2_{\rm{nors}}$ values are based off of optimizations to only the primary CMB anisotropy and BAO data. This was done to select parts of the parameter space preferred by these data to see if the CMB lensing, BBN, and supernova data strongly disfavored this cosmology; however, sampling all of these data together would result in a combination of the two that may result in less of an increase in $\Delta \chi^2_{\rm{nors}}$ as the fixed $H_0$ increases. This is illustrated in the $\chi^2_{\rm{rs}}-\chi^2_{\rm{nors}}$ panels of the triangle plots in Figure~\ref{fig:chi2_post} where there are parts of parameter space for EDE with $H_0 > 72$ \kmsmpc, $\chi^2_{\rm{rs}}$ lower than the \lcdm\ value (but not necessarily the minimum $\chi^2_{\rm{rs}}$ value overall), and comparable values of $\chi^2_{\rm{nors}}$.

\section{Conclusions} \label{Conclusions}

We constrained the \lcdm\ and EDE models with sound-horizon-independent data alone and in combination with sound-horizon-dependent data. We were motivated to perform these tests by new measurements from ACT DR6 CMB lensing and Pantheon+ supernova that constrained the Hubble constant because these data sets are independent of the size of the sound horizon. The main goal of this work was to determine what constraints the ACT DR6 CMB lensing data in combination with supernova data from Pantheon and Pantheon+ can place on the \lcdm\ and EDE models. 

For \lcdm, we found that these sound-horizon-independent data are sufficient to constrain $H_0$ precisely. Importantly, we found that the choice of supernova data, which is used to provide a constraint on $\Omega_m$ to break the $\Omega_m^{0.6}h$ degeneracy set by equality scale physics, is important in the determination of the resulting Hubble constant preferred value. Pantheon+ prefers a higher $\Omega_m$ than Pantheon and thus prefers a lower $H_0$ value. While the Pantheon+ preferred $\Omega_m$ is high even relative to \planck\ + BAO constraints \citep{planck/6:2018},  the latest supernova constraints from DES are also similarly high \citep{DES/Supernova}. We note that the constraining power of these data would be reduced if alternative dark energy, dark matter, or neutrino models are explored.

Unlike \lcdm, we found that the EDE phenomenology is largely unconstrained by the ACT and \planck\ CMB lensing, BBN, and supernova data, and importantly these unconstrained parameters allow for large increases in $H_0$ that can achieve $H_0$ values in agreement with the SH0ES measurement as well as the $H_0$ values preferred by the CMB + BAO data explored in this work when fit by EDE. We conclude that an extended cosmological model that modifies the size of the sound horizon at decoupling to increase the preferred value of $H_0$ can result in $H_0$ constraints from sound-horizon-dependent and independent data sets that are both consistent with each other and importantly agrees with the SH0ES $H_0$ value.

This conclusion is consistent with one of the main conclusions of \cite{Baxter_Sherwin/2021,Philcox/etal:2022,ACTDR6Madhavacheril}, which is that extensions to the standard model of cosmology attempting to resolve the $H_0$ tension by altering the size of the sound horizon must also consider changes to the physics at matter-radiation equality. In this case, the early dark energy scalar-field suppresses the growth of matter perturbations. Within EDE, an increase in the cold dark matter density can approximately offset this suppression; however, for general EDE-like models this increase in the cold dark matter density will not perfectly offset the changes made by the scalar-field and will require shifts in other cosmological parameters to optimally fit the sound-horizon-independent data. The prior on $\Omega_m$ from supernova data requires that any changes to $H_0$ will necessarily increase $\Omega_mh^2$ and thus $\Omega_ch^2$ given the BBN prior on $\Omega_bh^2$. Hence both the CMB lensing and supernova data allow for increases in $H_0$ within EDE if there are also sufficient increases in the cold dark matter density. 

Based on these results, we explored the constraining power of combining these data with primary CMB anisotropy and BAO measurements. The inclusion of CMB + BAO data restricts the allowed parameter space for EDE, which is important in determining whether EDE can optimally fit the sound-horizon-independent data even with less model freedom. We did this by taking \lcdm\ and EDE MCMCs constrained by ACT + \planck\ TT650TEEE + BAO and importance sampled them with the CMB lensing, BBN, and supernova data sets. We found that EDE is preferred by these combined data sets at approximately $2\sigma$ over \lcdm\ in both the cases where Pantheon or Pantheon+ is used, though the preference is weaker for Pantheon+. We found that most of this improvement comes from the improvement to the fit to primary CMB anisotropy measurements, which tends to increase as $H_0$ increases. In contrast, we found the fit to the various CMB lensing and supernova data is equivalent or worse for EDE compared to \lcdm, and tends to worsen as $H_0$ increases. 

For the CMB lensing and supernova data explored in this work, we tested whether the worsening of the fit to these data was statistically significant given a prior set by the constraint from the primary CMB anisotropy and BAO data. We found that while the EDE model fits these data sets worse than \lcdm\ when restricted to the allowed parameter space by ACT + \planck\ TT650TEEE + BAO measurements, the fit to the data is still consistent to within 1$\sigma$ of being a random statistical fluctuation. We conclude from these tests that EDE is still allowed by these data. Moreover, we conclude that assuming the late universe phenomenology is close to \lcdm, constraints from uncalibrated supernova will play an important role in constraining EDE as high preferred values of $\Omega_m$ reduce the allowed EDE parameter space with $H_0$ values above 70 \kmsmpc.

We caution that these results do not imply that EDE is the correct cosmological model or that there must be a miscalibration of the size of the sound horizon. While EDE does fit the ACT + \planck\ TT650TEEE + BAO better than \lcdm, it fits the CMB lensing and Pantheon+ data worse than \lcdm. While this worsening of the fit was found not to be statistically significant, EDE has been shown to generally fit LSS data worse than \lcdm. Moreover, the preference for EDE over \lcdm\ when fitting the ACT + \planck\ TT650TEEE + BAO data may result from noisier measurements that may go away with future CMB data. Finally, there is still no preference within a Bayesian perspective for EDE when \planck\ TT $\ell > 650$ data are also included. All of these problems suggest that EDE, in its simplest form, is not likely to be the next concordance model of cosmology. 

Nevertheless, we highlight that while this analysis uses EDE as a specific model that modifies the size of the sound horizon, the results of this work are more general. We conclude that other models that rely on a miscalibration of the sound horizon could also be allowed by the sound-horizon-independent data explored in this work. Moreover, we highlight that a cosmological model that relies on a miscalibration of the sound horizon to resolve the Hubble tension could be designed to more optimally offset the changes to the growth of matter perturbation resulting from increasing the cold dark matter density. We note that we have not used all possible current or possible future measurements that are independent of the sound horizon. We leave these explorations to future work.  

\acknowledgments

We would like to thank Graeme Addison, Neelima Sehgal, and Tanvi Karwal for their helpful suggestions and discussions. We would also like to thank Stony Brook Research Computing and Cyberinfrastructure, and the Institute for Advanced Computational Science at Stony Brook University for access to the high-performance SeaWulf computing system, which was made possible by National Science Foundation grant (1531492). We acknowledge the use of the Legacy Archive for Microwave Background Data Analysis (LAMBDA), part of the High Energy Astrophysics Science Archive Center (HEASARC). HEASARC/LAMBDA is a service of the Astrophysics Science Division at the NASA Goddard Space Flight Center.

\bibliographystyle{JHEP}
\bibliography{cosmology}

\end{document}